\documentclass[10pt]{article}
\usepackage{psfig,epsf,conf-X}
\def\approxgt{\mathrel{\hbox{\rlap{\lower.55ex \hbox {$\sim$}}
        \kern-.3em \raise.4ex \hbox{$>$}}}}
\def\approxlt{\mathrel{\hbox{\rlap{\lower.55ex \hbox {$\sim$}}
        \kern-.3em \raise.4ex \hbox{$<$}}}}
\begin{document} 
\small

\heading{XEUS - The X-ray Evolving Universe Spectroscopy Mission}
\author{A.N. Parmar$^1$, T. Peacock$^1$, M. Bavdaz$^1$, G. Hasinger$^2$,
 M. Arnaud$^3$, X. Barcons$^4$, D. Barret$^5$, A. Blanchard$^6$,
 H. B\"ohringer$^7$, M. Cappi$^8$, A.~Comastri$^9$,
 T. Courvoisier$^{10}$, A.C. Fabian$^{11}$, R. Griffiths$^{12}$,
 P.~Malaguti$^8$, K.O. Mason$^{13}$, T. Ohashi$^{14}$, 
 F. Paerels$^{15}$,
 L. Piro$^{16}$, J. Schmitt$^{17}$, M.~van der Klis$^{18}$, 
 M. Ward$^{19}$}

\address{Space Science Department of ESA, ESTEC, NL}
\address{Astrophysikalisches Institut Potsdam, 14482 Potsdam, Germany}
\address{CEA/DSM/DAPNIA/SAP CEN-Saclay, Gif-sur-Yvette, France}
\address{IFCA Universidad de Cantabria (CSIC-UC) Santander, Spain}
\address{CESR-CNRS/UPS, Toulouse, France}
\address{Observatoire Astronomique de Strasbourg, Strasbourg, France}
\address{MPE f\"ur Extraterrestrische Physik, Garching bei M\"unchen, Germany}
\address{Istituto TESRE, CNR, Bologna, Italy}
\address{Osservatorio Astronomico di Bologna, 40127 Bologna, Italy}
\address{Observatoire de Geneve, Sauverny, Switzerland}
\address{Institute of Astronomy, Cambridge, UK}
\address{Carnegie Mellon University, Pittsburgh, PA, USA}
\address{Mullard Space Science Laboratory, UCL, Dorking, UK}
\address{Tokyo Metropolitan University, Hachioji, Tokyo, Japan}
\address{Space Research Organization Netherlands, Utrecht, NL}
\address{Istituto di Astrofisica Spaziale (IAS), CNR, 00133 Rome, Italy}
\address{Universit\"at Hamburg, Hamburger Sternwarte, Hamburg, Germany}
\address{Astronomical Institute Anton Pannekoek, University of Amsterdam, NL}
\address{X-ray Astronomy Group, University of Leicester, Leicester, UK}

\begin{abstract}
XEUS is under study by ESA as part of the Horizon 2000+ program
to utilize the International Space Station (ISS) for astronomical 
applications. XEUS will be a long-term X-ray observatory with
an initial mirror area of 6~m$^2$ at 1~keV that will be grown to
30 m$^2$ following a visit to the ISS.
The 1~keV spatial resolution is expected to be 2--5$''$ HEW. 
XEUS will consist of separate detector and mirror spacecraft aligned by 
active control to provide a focal length of 50~m. 
A new detector spacecraft, complete with the next generation of
instruments, will also be added after visiting the ISS. The
limiting sensitivity will then be
$\sim$$4 \times 10^{-18}$~erg~cm$^{-2}$~s$^{-1}$, 
around 250 times better than XMM, allowing XEUS to study the properties of 
the hot baryons and dark matter at high redshift.

\end{abstract}

\section{Introduction}

XEUS, the X-ray Evolving Universe Spectroscopy mission, is a potential 
follow-on mission to XMM and is being studied as part of the Horizon 2000+ 
program within 
the context of the International Space Station (ISS) utilization.  
The XEUS mission aims to place a long lived X-ray observatory in space with a 
sensitivity comparable to the next generation of ground and
space based observatories such as ALMA and NGST (Fig.~1).
By making full use of the facilities available 
at the ISS and by ensuring in the design a significant 
growth and evolution potential, the overall mission lifetime of XEUS could be 
$>$25 years. 

\begin{figure}
       \hbox{\hspace{2.0cm}
       \psfig{figure=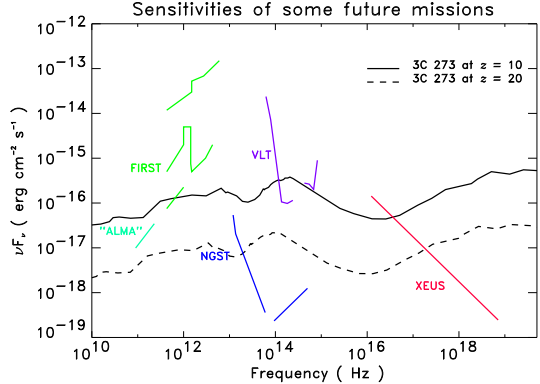,width=7.5cm,angle=-0}}
      \caption[]{\small Comparison of the sensitivities of future missions in
                 different wavebands. A horizontal line corresponds to 
                 equal power output per decade of frequency. For ALMA
                 an 8~hr integration was assumed, for FIRST a 5$\sigma$
                 detection in 1~hr, for NGST a 5$\sigma$ detection in 10~ks,
                 and for XEUS a 100~ks exposure}
\end{figure}

The key characteristic of XEUS is the large aperture X-ray mirror. 
This will capitalize on the successful XMM mirror 
technology and the industrial foundations which have been already laid in 
Europe for this program.
The XEUS 
mirror aperture of 10~m diameter will be divided into annuli with each 
annulus 
sub-divided into sectors. The basic mirror unit therefore consists of a set 
of heavily stacked thin mirror plates. 
This unit is known as a ``mirror 
petal'' and is a complete, free standing, calibrated part of the overall 
XEUS optics with a spatial resolution of 2--5$''$ HEW and a broad energy 
range of 0.05--30~keV. Each mirror petal will be 
individually alignable in orbit.
Narrow and Wide field imagers will provide
FOVs of 1$'$ and 5$'$, 
and energy resolutions of 1--2~eV and 50~eV at 1~keV. 

\section{Mission Profile}

XEUS will consist of separate detector 
(DSC) and mirror spacecraft (MSC) separated by 50~m and 
aligned by active control.
The large aperture mirror cannot be deployed in a 
single launch. Instead, 
the ``zero growth'' XEUS (MSC1+DSC1) will be 
launched directly into a Fellow Traveler 
Orbit (FTO) to the ISS using an Ariane~V or similar. The FTO is a 
low Earth orbit with an altitude of $\sim$600~km and 
an inclination similar to the ISS. 
The mated pair will then decouple and DSC1 will take up 
station 50~m from the MSC1 and after check-out 
the zero growth astrophysics observation 
program will commence with an aperture of 6~m$^2$ at 1~keV. 

After 4--5 years of observations, the XEUS spacecraft will 
re-mate and maneuver to the vicinity of the ISS. 
At the ISS the MSC1 will separate from DSC1 
and then dock with the ISS. 
The DSC1, with 
its usefulness at an 
end, will undergo a controlled de-orbit.
At the ISS the mirror area is expanded to 30~m$^2$ at 1~keV (see Fig.~2) 
and MSC1 becomes 
MSC2. The extra mirror petals will have already been transported
to the ISS using the STS or the European Automated Transfer 
Vehicle (ATV). Once the
mirror growth and checkout is complete, MSC2 will leave the
ISS and mate with the recently launched DSC2. Using the DSC2
propulsion system the pair will return to FTO
and the fully grown XEUS will start its observing program.

\begin{figure}
       \hbox{\hspace{2.0cm}
       \psfig{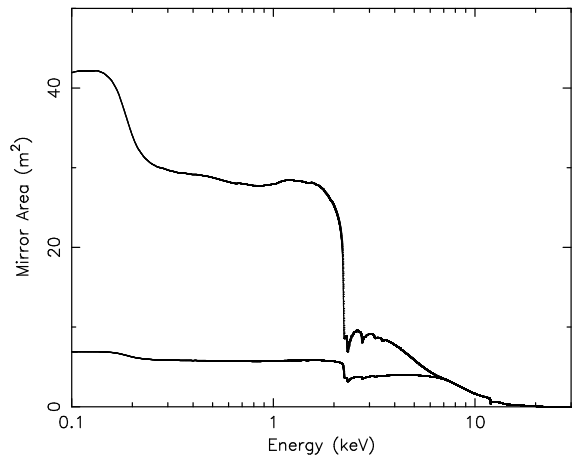}}
      \caption[]{\small The XEUS mirror area before and after growth at the 
                 ISS}
\end{figure}

\section{Science Goals}

XEUS will study the evolution of the hot baryons in the Universe and
in particular:

\begin{itemize}
\item{} Detect massive black holes in the earliest AGN and estimate their
mass, spin and $z$ through studies of relativistically broadened Fe-K lines
and variability.
\item{} Study the formation of the first gravitationally bound, 
dark matter dominated, systems ie. small groups of galaxies and 
trace their evolution into today's massive clusters. 
\item{} Study the evolution of metal synthesis down to the present epoch,
using in particular, observations of the hot intra-cluster gas.
\item{} Characterize the mass, temperature, density of the intergalactic
medium, much of which may be in hot filamentary structures, using 
aborption line spectroscopy. High $z$ luminous quasars and X-ray afterglows
of gamma-ray bursts can be used as background sources.
\end{itemize}

\subsection{Spectroscopy of Massive Black holes}

\begin{figure}\hbox{
       \psfig{figure=parmarF3a.ps,width=5.6cm,height=3.0cm,angle=-90}
      \hspace{0.0cm}
      \psfig{figure=parmarF3b.ps,width=5.6cm,height=3.0cm,angle=-90}}
      \hbox{
      \psfig{figure=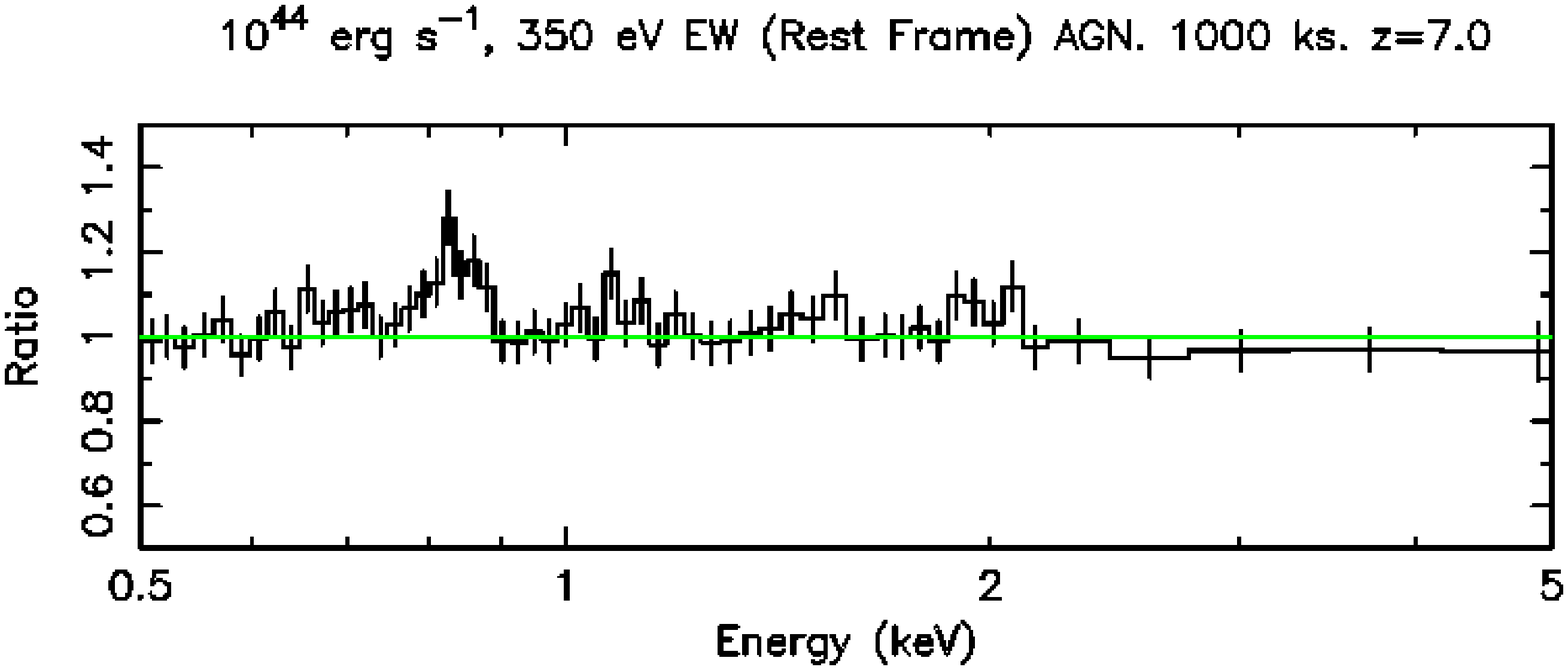,width=5.6cm,height=3.0cm,angle=-0}
      \hspace{0.0cm}
      \psfig{figure=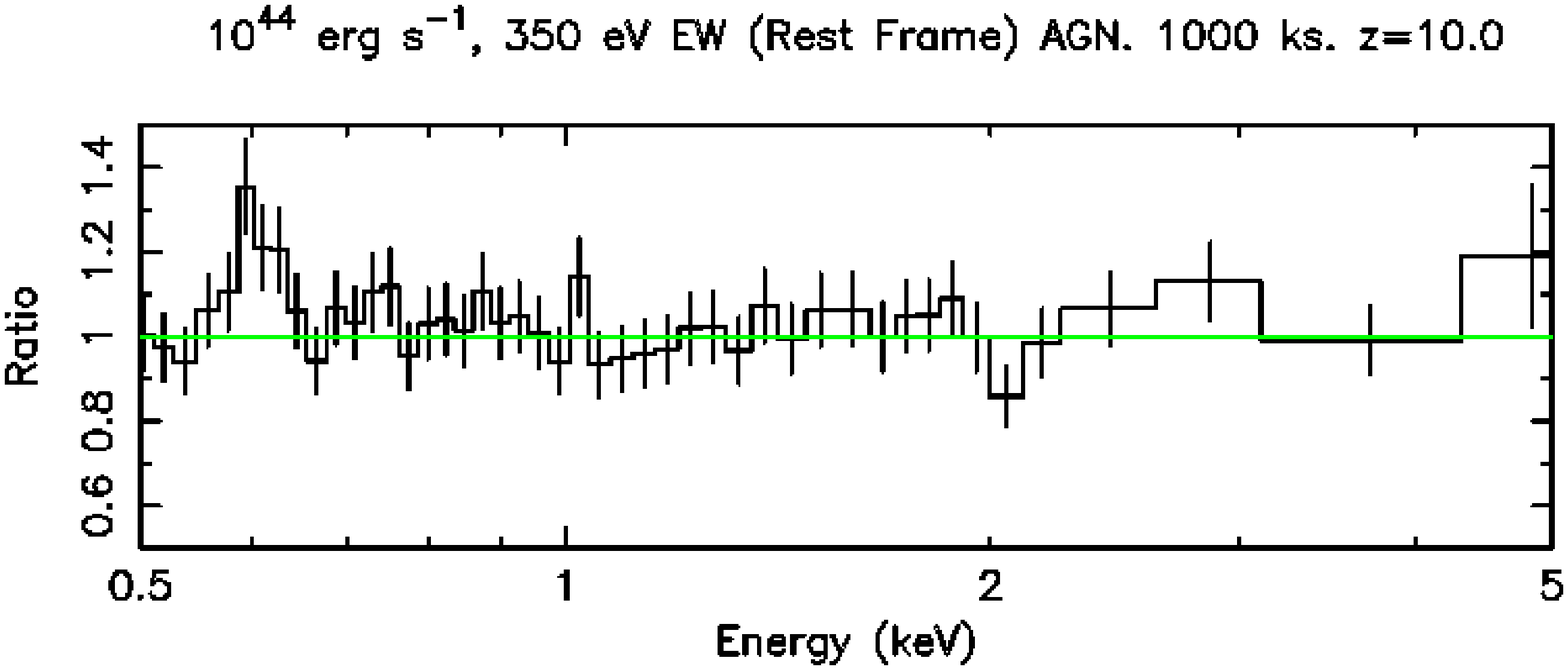,width=5.6cm,height=3.0cm,angle=-0}}
      \caption[]{\small The residuals when the 350~eV EW Fe line
      normalization is set to zero for an AGN with a 
      (rest-frame) luminosity of $10^{44}$~erg~s$^{-1}$ for $z$ = 3, 5,
      7, and 10 for the fully grown XEUS configuration.}
\end{figure}

Currently, X-ray astronomy can only detect AGN 
to a $z$ of $\sim$5. XEUS will be able to
undertake {\it detailed} X-ray spectroscopy of much more distant AGN. 
Fig.~3 illustrates the results of a 
series of simulations of a ``typical'' AGN
with a 2--10~keV rest-frame luminosity of $10^{44}$~erg~s$^{-1}$ at
different red-shifts. An
exposure time of 10$^6$~s was assumed for the fully grown XEUS.
Values for H${\rm _0}$ and q${\rm _0}$ of 
50~km~s$^{-1}$~Mpc$^{-1}$ and 0.5 together with
an underlying ${\rm E^{-2.0}}$ spectrum with
a Galactic ${\rm N_H}$ of
$10^{21}$~atom~cm$^{-2}$ and a local (red-shifted)
${\rm N_H}$ of $5 \times 10^{21}$~atom~cm$^{-2}$ were assumed. 
A ``double-horned'' relativistically distorted and Doppler broadened
Fe line at 6.4~keV with a rest-frame equivalent width of 350~eV
was simulated. The other line parameters were taken to be 
as for MCG-6-30-15.
Fig.~3 shows the residuals when the source is
red-shifted to $z$ = 3, 5, 7, and 10, demonstrating that such a line
can be clearly detected and its properties measured even at $z$ = 10.

\begin{figure}
       \hbox{\hspace{2.0cm}
       \psfig{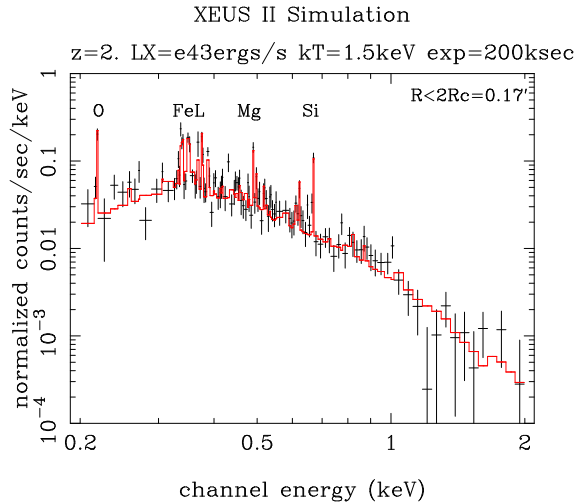}}
      \caption[]{\small Simulated XEUS spectrum of a $z$ = 2 galaxy group.}
\end{figure}

\subsection{Spectroscopy of Distant Galaxy Groups}

To illustrate the potential of XEUS to study the formation of large
scale structure Fig.~4 shows a simulation of a distant ($z$ = 2)
galaxy group. In standard
cosmological models these groups are the first emerging massive objects,
with masses of $\sim$$10^{13}$M$_{\odot}$. The epoch of their first 
formation depends critically on the adopted cosmology, and is
likely to be $z \sim 2$--5. Therefore the study of groups will 
provide a deep probe of the early Universe. These systems and their
dark matter aloes are the smallest units by which to study the hot
thermal intergalactic gas trapped in deep gravitational wells. 
Emission lines of O, Fe, Mg, and Si are clearly evident.
The temperature can be determined to better than $\pm$3\% and the Fe
and O abundances to better than 10\% and 20\%, respectively.

\subsection{Resonant Absorption Line Studies}

XEUS will be the first X-ray
observatory capable of detecting resonance absorption lines for a wide
range of objects.  This results from the unique combination of 
large effective area and high spectral and spatial resolutions.
The use of resonance absorption lines can be applied to several
problems, as it is in optical/UV astronomy. Resonance absorption
lines are generally detectable at much lower column densities than
absorption edges (which do not require high resolution spectroscopy),
and therefore can trace gas which is too tenuous to be seen by other
means.

Intervening hot/warm gas clouds along the line of sight towards
distant background sources will produce resonance absorption
lines. The main
issues that can be addressed with these studies
include the use of absorber number counts and their redshift dependence to
test models of large-scale structure formation, the
determination of the temperature distribution of baryons in the Universe,
the determination of metallicities of the absorbers, and in particular
the [O/Fe] ratio to infer the relative rates of type I and II Supernovae,
the determination of the redshift evolution of parameters such as
number counts, gas kTs, and metallicities, and
when the emitting gas is also seen in absorption, the use
both emission and absorption to infer distances, and therefore
measure key cosmological parameters.

\begin{figure}
\centerline{\psfig{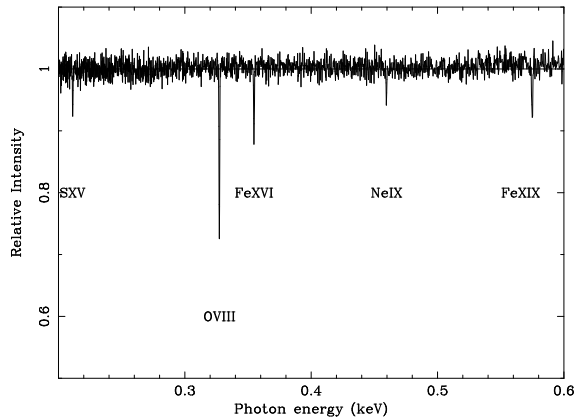}}
\caption{\small Absorption line spectrum towards a ${\rm 10^{-12}\, erg\,
cm^{-2}\,  s^{-1}}$ AGN (0.5--2 keV) observed for 100 ks
with XEUS2.  The line of sight is assumed to cross a small group
(L=$10^{42}{\rm erg}\, {\rm s}^{-1}$) at $z=1$ with a core of
radius 50 kpc. For a $10^{-13}\, {\rm erg}\, {\rm cm}^{-2}\, {\rm
s}^{-1}$ background source, the O~{\sc viii} line is still clearly
detected.}
\end{figure}

\subsection{Studying Dust Enshrouded AGN and Starburst Galaxies}

\begin{figure}
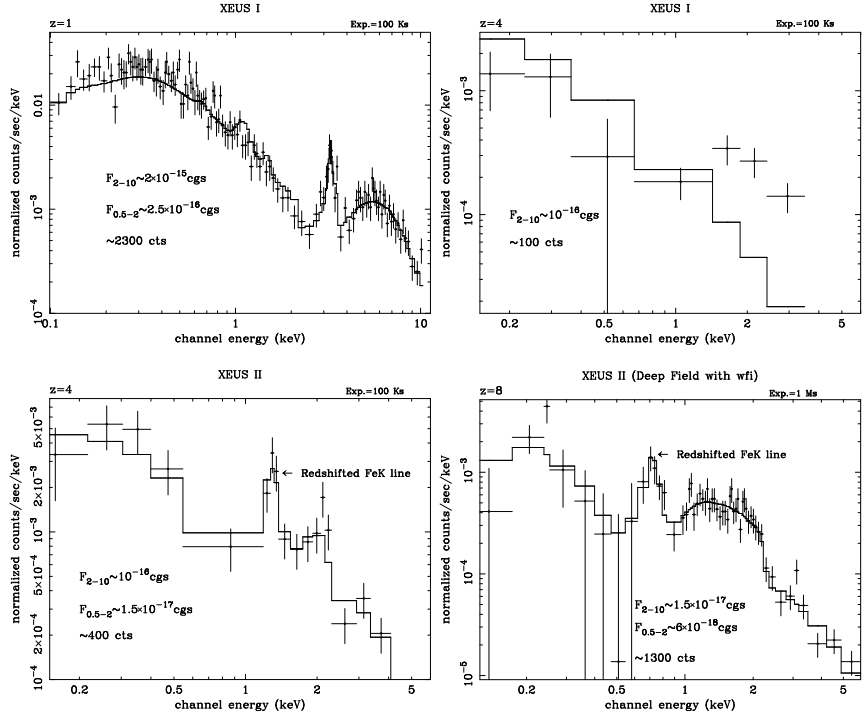

\hbox{
\psfig{file=parmarF6a.ps,width=8cm,width=5.6cm,angle=-90}
\psfig{file=parmarF6b.ps,width=8cm,width=5.6cm,angle=-90}}
\vspace{0.3cm}
\hbox{
\psfig{file=parmarF6c.ps,width=8cm,width=5.6cm,angle=-90}
\psfig{file=parmarF6d.ps,width=8cm,width=5.6cm,angle=-90}}
\caption{\small Upper panels: simulated XEUS1 spectra at $z=1$ (Left) and 
$z=4$ (Right). These illustrate the limits of the XEUS prior to
growth at the ISS. The lower panels show simulated XEUS2 spectra (after
mirror growth) at $z=4$ (Left) and $z=8$ (Right).}
\label{SIM1}
\end{figure}

In order to test the sensitivity of XEUS 
to discriminate between AGN and starburst emission,
spectra of a composite starburst 
galaxy plus a heavily absorbed AGN have been simulated. 
The starburst emission was parameterized by a thermal gas at kT = 3 keV with 
0.3 solar metallicity.
Above a few keV the absorbed AGN is expected to show up with a strong 
(EW = 1 keV) Fe-K line 
due to transmission through the N$_{\rm H}$=10$^{24}~$cm$^{-2}$ 
absorbing material.
Such a model is similar to 
that of 
the nearby galaxies NGC~6240, NGC~4945, and Mkn~3. 
Fig. \ref{SIM1} demonstrates that XEUS1 will allow 
a detailed study 
of such sources around $z=1$, but that XEUS2 is required to perform 
spectroscopy 
at $z\gg1$. 
Such X-ray spectra are the only way to obtain {\it direct} proof of the 
existence of dust-enshrouded AGNs at high redshift. 
If detected, they would allow
the starburst versus AGN contribution to be directly disentangled. If 
undetected, 
they would give strong limits on the AGN contribution. This would have 
important consequences on the star formation and ionization histories of 
the Universe.

\subsection{Stellar Spectroscopy}

The large effective area of the XEUS configuration provides unique
opportunities for stellar X-ray astronomy.  
The high sensitivity means that solar-like X-ray emission can be detected
out to distances of a few kpc. As a consequence large samples of 
truly solar-like stars
become amenable for study.  For example, at the distance of M~67, an
old open cluster with an age similar to that of the Sun, a
limiting X-ray luminosity of 10$^{26}$ erg~s$^{-1}$ can be reached, implying
that solar minimum X-ray emission levels can be detected.  This is 
particularly relevant for a study of activity cycles in other solar-like
stars, since in the Sun the solar cycle is most easily detectable in the
X-ray domain.  In addition, the XEUS
sensitivity is so large that in nearby open clusters such as Hyades and
Pleiades virtually all cluster stars will be detectable as X-ray sources,
and the X-ray brightest cool stars can even be detected in nearby galaxies
such as the LMC and M~31.

\begin{figure}[htp]
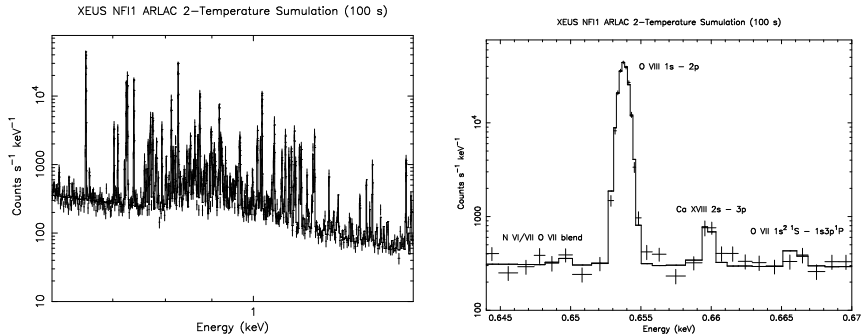

\hbox{\psfig{figure=parmarF7a.ps,width=5.4cm,angle=-90}
      \hspace{0.3cm}
      \psfig{figure=parmarF7b.ps,width=5.4cm,angle=-90}}
\caption{\small 
(left) Parts of a simulated NFI1 spectrum of AR Lac with the XEUS1 
configuration showing details of the Fe-L region (left). 
An exposure time of 100 s was used. A similar quality spectrum is expected in 
only 20 s with XEUS2. (right) The spectral region near oxygen Ly-$\alpha$ 
with line identifications.} 
\label{ARL}
\end{figure}
 
X-ray images of the Sun have revealed that hot plasma trapped in closed
magnetic loops provides almost all of the solar X-ray emission.
While such X-ray emission is usually ``quiet'', sometimes
restructuring of such magnetic loops gives rise to
intense outbursts of radiation in the form of flares.  On other stars
flare much more intense than those on the Sun are observed.
Time resolved high resolution
spectroscopy is required to understand and analyze the physics of such
giant stellar flares. The potential of XEUS to perform such studies 
is illustrated by simulations of the
nearby RS CVn system AR Lac (G2 {\sc iv} + K0 {\sc iv}). 
The 100~s simulations shown in Fig.
\ref{ARL} show parts of a rich line-dominated spectrum. 
The large area of XEUS means that a
sufficient number of counts are obtained so that the temperature, density, 
chemical abundance and velocity distribution of the emitting plasma can 
be measured on very short timescales. This will allow the study of
the evolution
of these basic physical parameters during typical stellar flares with 
an accuracy only previously achievable with solar flares.

\acknowledgements{
We thank the XEUS Steering Committee (M. Turner, J. Bleeker, G. Hasinger
H. Inoue, G. Palumbo, T. Peacock and J. Tr\"umper) and the ESA ISS and XMM 
project teams for their support.}

\end{document}